\documentclass[aps,prl,twocolumn,groupedaddress]{revtex4}
\usepackage{graphicx,subfigure}
\begin{document}
\title{Metal-Semiconductor Transition in Armchair Carbon Nanotubes\\
 by Symmetry Breaking}

\author{Yan Li}
\author{Slava V.~Rotkin} 
\thanks{Present address: Department of Physics, Lehigh University,
16 Memorial Drive East, Bethlehem PA 18015, USA.}
\author{Umberto Ravaioli}
\affiliation{Beckman Institute for Advanced Science and Technology, University of Illinois at Urbana Champaign, 405 N. Mathews, Urbana, Illinois 61801}
\date{\today}

\begin{abstract}
The electronic band structure of armchair carbon nanotubes may be
considerably modified by potentials with angular dependence. Different
angular modes $V_q\sim \cos q \theta$ have been studied within a
tight-binding scheme. Using symmetry arguments, we demonstrate a
band gap opening in these metallic nanotubes when certain
selection rules are satisfied for both potential and nanotube
structure. We estimate the band gap opening as a function of both
the external potential strength and the nanotube radius and
suggest an effective mechanism of metal-semiconductor transition
by combination of different forms of perturbations.
\end{abstract}
\maketitle

Armchair single-wall carbon nanotubes (SWNTs) with indices $(n,n)$
are metallic with two subbands crossing at the Fermi level, which
is allowed by opposite parities of these two subbands with regard
to the $n$ vertical mirror reflections $\sigma_v$. It is
interesting to explore if one can break the mirror symmetry and
generate a band gap by choosing appropriate angular perturbations,
so that one can modify and control nanotube material properties
for applications. It was recently proposed to use a very
inhomogeneous electric field to induce metal-semiconductor
transition (MST) in armchair SWNTs~\cite{ROTK2004}. Other types of
circumferential perturbations were studied including intra-rope
interaction \cite{DELA98}, twisting or bending \cite{KANE97},
squashing \cite{PARK99,LAMM2000,GULS2002a,LU2003} and applying
uniform perpendicular electric fields \cite{OKEE2002,LI2003}.  Some
perturbations indeed were found to induce MST, which was
attributed partially to the breaking of SWNT mirror symmetry.
However, there still remains an open question if mirror symmetry
breaking (MSB) is a sufficient condition to bring in a band gap in
armchair tubes. For example, in the case of bending or applying a
uniform electric field, by rotating the armchair nanotube, one can
always find an alignment that breaks all vertical mirror
symmetries but the nanotube remains metallic, which cannot be
explained by the MSB argument alone. 

In this letter, we apply potentials with angular dependence to
modulate the band structure and electronic properties of the
armchair SWNTs. We choose angular modes $V_q=V_0\cos{q\theta}$ or
their combinations to study gap opening within an orthogonal
$\pi$-orbital tight-binding (TB) scheme.  We found that (1) breaking
symmetries about all the vertical mirror reflections and
$C_2$ rotations (see below) is required to mix the two
linear subbands near the Fermi level; (2) selection rules of
subband coupling impose additional requirement on the angular
momentum of the perturbation, e.g. a single mode with odd $q$ does
not generate any band gap; (3) for modes with even $q$, MST is
only possible for tubes with specific indices $n$ to satisfy
selection rules described below.

\begin{figure}[b]
\includegraphics[angle=0,width=3.375 in]{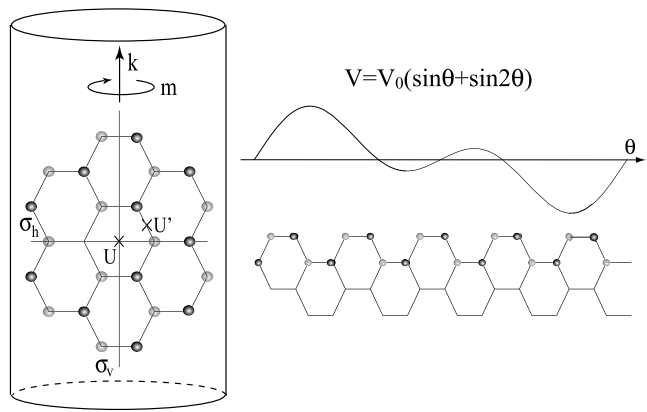}
\caption{{\bf Left}: Symmetry operations of an $(n,n)$ armchair
nanotube. {\bf Right}: Unwrapped unit cell of a (5,5) nanotube and
schematics of an external potential of the form
$V=V_0(\sin\theta+\sin2\theta)$. \label{fig1}}
\end{figure}

\emph{Selection Rules.}  The symmetry group of an $(n,n)$ armchair
SWNTs consists of vertical mirror planes $\sigma_v$, horizontal
mirror planes $\sigma_h$ and rotation axes
$C_2\equiv\{U,U'\}$~(Fig.~\ref{fig1}) . Every electron state
can be labeled by a set of quantum numbers: an axial wave number
$k$, an angular quantum number $m$, a parity with regard to the
vertical mirror reflection $\sigma_v (A/B)$ and a parity to the
horizontal mirror reflection $\sigma_h (+/-)$.  There are also two
sets of $C_2$ axes $U$ and $U'$, which are perpendicular to the tube
axis. The two linear subbands $|\pi\rangle$ and $|\pi^*\rangle$, both
have $m=n$ and even parities about $\sigma_h$, while their parities
about $U,U'$ and $\sigma_v$ are opposite~\cite{VUKO2002}.  It is the
different \emph{parities} that allow the two subbands to cross and be
degenerate at the Fermi points. We do not consider potentials with an
axial space dependence, thus, the quantum number $k$ is conserved.

What are the symmetry conditions required for MST in armchair
nanotube?
\begin{description}\label{rule1}
  \item[ Rule 1] All symmetries about vertical mirror planes $\sigma_v$
  and $C_2$ axes $U,U'$ must be broken simultaneously.
\end{description}

\begin{figure}[b]
\includegraphics[width=3.375 in]{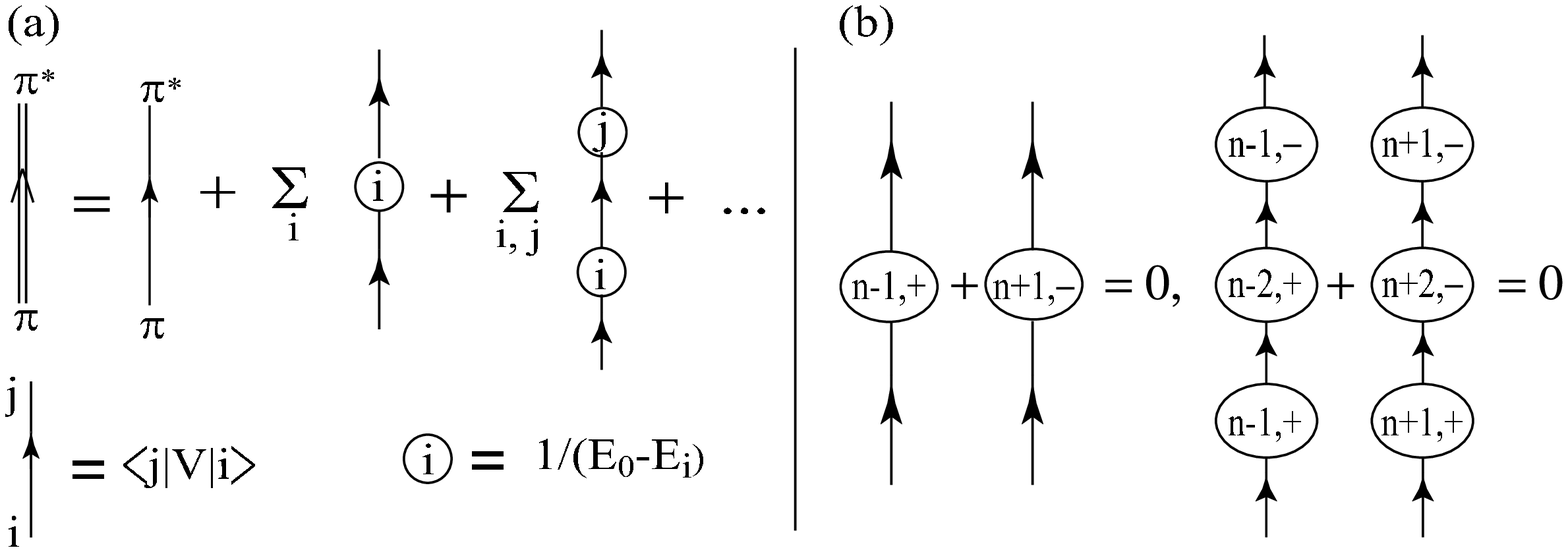}
\caption{(a) Feynman diagram for interaction between $|\pi\rangle$
and $|\pi^*\rangle$ states. $|i\rangle=|m_i,s_i\rangle$. (b)Two
examples of the contributions that cancel out for $\mu=\mbox{even}$.\label{fig2}}
\end{figure}

The selection rules of subband coupling impose additional requirement
on the \emph{angular quantum number} of the potential as well as the
nanotube indices. Assume that the external potential has only one
angular mode $V_q(\theta)\sim\cos q(\theta+\theta_0)$, where an
arbitrary offset $\theta_0$ is zero when a mirror plane of the
potential coincides with the one of the SWNT. The conservation of
total quantum numbers $k$ and $m$ imposes the selection rules for
direct subband coupling in an $(n,n)$ SWNT: $\delta k=0$ and $\delta
m=\pm q+2nj$, with $j$ an integer. The indirect interaction between
$|\pi\rangle$ and $|\pi^*\rangle$ states $H_{\pi\pi^*}$ can thus be
represented by a Feynman-like diagram, or as a perturbation series of
the coupling order $\mu$ (Fig.~\ref{fig2}(a)). Here all
allowed intermediate states $\{m_i,s_i\}$ are states of the given
angular momentum $m$ and pseudo-spin $s$, with $s=\pm1$ denoting the
conduction and valence bands respectively. Due to the symmetry of
electron and hole bands, contributions of even $\mu$ cancel out as
shown in Fig.~\ref{fig2}(b) (a detailed example will be shown
below). Thus, only for odd coupling order $\mu$, there can be a
non-vanishing coupling between $|\pi\rangle$ and $|\pi^*\rangle$
states. The most important contribution comes from the \emph{lowest
possible} $\mu$ which is $\mu_0=2n/\mathrm{gcd}(2n,q)$, where
$\mathrm{gcd}$ is the greatest common divisor. We formulate the second
rule for the MST in the single mode angular potential $V_q$:
\begin{description}\label{rule 2}
  \item[Rule 2]\[\mu_0\equiv\frac{2n}{\mathrm{gcd}(2n,q)}=\mbox{odd}.\]
\end{description}

 One conclusion we can draw immediately is that there is no MST when
 $q$ is odd, because then $\mu_0$ is even, which
 violates rule 2. This is consistent with the absence of the band gap for
 armchair SWNTs under a uniform perpendicular electric field,
 i.e. $q=\pm1$ \cite{LI2003}.
 
By nearly degenerate perturbation theory, we derive the
dependence of the band gap opening on the angular offset
$\theta_0$. We already know that when $\theta_0=0$, there shall be no
band gap according to rule 1. By changing the relative alignment of
the SWNT and the potential, a band gap starts to develop and $E_g$ can
be estimated as:
\begin{equation}
  E_g(\theta_0)=E_g^{max}\sin\left(\mu_0q\theta_0\right).
\end{equation}
It is self-evident that the maximum of the band gap is for
$\theta_0=\pi/(2\mu_0q)$ when $V_q$ is so aligned that
the potential is antisymmetric about one of the vertical mirror
planes.

 The case with $q=2$ is reminiscent of squashing an armchair SWNT. Our
 analysis explains why a small band gap was found in a $(5,5)$ SWNT in
 Ref.~\cite{PARK99} but not in $(6,6)$ SWNT in Ref.~\cite{GULS2002a}
 or $(8,8)$ SWNT in Ref.~\cite{LU2003}. Since $\mu_0=n$ for $q=2$,
 only SWNTs with odd $n$ can open a band gap. In (8,8) and (6,6)
 armchair nanotubes there should be no mixing between the crossing
 subbands. For odd $n$, the band gap opening is roughly
$E_g\propto
V_0^n\sin(2n\theta_0)$ and decreasing for large radius
 nanotubes due to the higher order of perturbation. The maximum band gap of a (5,5) SWNT by the $q=2$
 potential is of the order of $0.1$ eV within our TB calculation.

 The most interesting case for potential applications is that the
 potential has short oscillation period: $q=2n$, which yields direct
 mixing to the coupling order $\mu_0=1$. Assuming no overlap between
 orbitals on different atomic sites, an analytical expression for the
 band gap can be obtained by the degenerate perturbation theory:
 \begin{equation}\label{q=2n}
   E_g=\sqrt{3}V_0\sin(2n\theta_0).
 \end{equation}
Since $|\pi\rangle$ and $|\pi^*\rangle$ are now directly coupled,
the band gap is proportional to the perturbation and the relation
in Eq. (\ref{q=2n}) holds up to a few eV.  Potential of this form
($q=2n$) requires changing the sign of the electrostatic potential
alternatively on neighboring carbon atoms.  One can possibly
generate such perturbations by twisting, chemical/biological
decoration of the tube or by using high multipoles of very
inhomogeneous potential.

\emph{The potential with mixed Fourier components.} Realistic
perturbations usually have more than one dominating angular mode,
which makes analytical expressions tedious. On the other hand, the
interplay of the different angular components may result in
stronger and more interesting perturbation of the electronic
properties of the SWNTs. For example, the transport behavior of a
deformed nanotube may change under the control of the gate
potential~\cite{BIER2003} . When the strength of the field across
the SWNT is large, the gate potential shifts the Fermi
level and also modifies the band structure. In the simplest case, only
two angular components are present:
$H_1=V_1\cos(\theta+\theta_1)+V_2\cos\left(2(\theta+\theta_2)\right)$.
Here $\theta_1$ and $\theta_2$ are defined as the angular offsets
of the mirror planes of $V_1$ and $V_2$ with regard to that of the
SWNT.

The coupling order $\mu_0$ is now a function of both $q_1$ and
$q_2$. For $q_1=1,q_2=2$, the lowest allowed coupling is of the $3$rd
order through $|\pi\rangle\leftrightarrow
|m_1,s_1\rangle\leftrightarrow
|m_2,s_2\rangle\leftrightarrow|\pi^*\rangle$ and there are 24 possible
sets of intermediate states $\{m_1,s_1;m_2,s_2\}$:
\begin{equation}\label{allowed transition}
  (m_1,m_2)=\left\{\begin{array}{ccc}
(\pm1,\mp1)\\
(\pm1,\pm2)\\
(\pm2,\pm1)
\end{array}\right.\otimes\;\;
(s_1,s_2)=\left\{
\begin{array}{ccc}
  (\pm1,\pm1) \\
  (\pm1,\mp1)
\end{array}\right..
\end{equation}
Here $m_{1,2}=\pm1,\pm2$ are the quantum numbers relative to
$m=n$. In general case, the band gap opening is related to the
off-diagonal element $H_{\pi\pi^*}$ in the $2\times 2$ matrix of
$\pi$ and $\pi^*$ states, which can be written via contributions
from different intermediate sets of
$|i\rangle\equiv|m_i,s_i\rangle $:
\begin{eqnarray}\label{off-diagonal}
H^{(\mu)}_{\pi\pi^*}(\{i\})&=&\frac{\langle \pi|H_1|1\rangle\ldots\langle i|H_1|i+1\rangle \ldots \langle \mu-1|H_1|\pi^*\rangle}{\left(E_0-E(1)\right)\ldots\left(E_0-E(\mu-1)\right)},\nonumber\\
  H_{\pi\pi^*}&=&\sum_{\{m_i,s_i\}}H^{(\mu)}_{\pi\pi^*}(\{i\}),
\end{eqnarray}
where $E_0=0$ for symmetric $\pi$ and $\pi^*$ bands and $E(i)$ is
the eigen energy of the electron state $|i\rangle$.

\begin{figure}[b]
\includegraphics[angle=0,width=3 in]{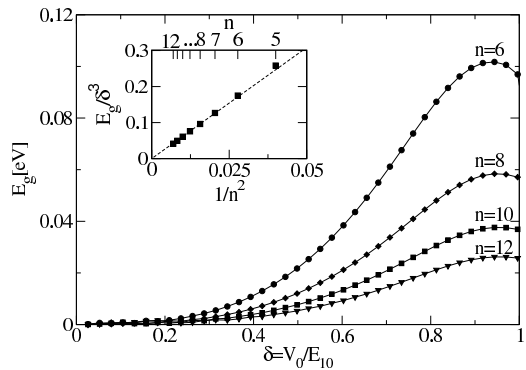}
\caption{The band gap opening as a function of the dimensionless
potential $\delta=V_0/E_{10}$ for $(n,n)$ SWNTs with $n=6, 8, 10, 12$, and
$V_1=V_2\equiv V_0$. Inset: The ratio $E_g/\delta^3$ vs.~$n^{-2}$ for TB
results (solid squares) and results of perturbation theory (dashed
line) with $n=5, \ldots , 12$.
\label{fig3}}
\end{figure}

The matrix element $\langle V \rangle=\langle i|V_q\cos
q(\theta+\theta_q)|i+1\rangle$ can be simplified by neglecting the
coupling between sub-lattices in an envelope function
approximation:
\begin{eqnarray}\label{matrix_element}
\langle V\rangle&\approx&\delta_{\Delta m,\pm q}\frac{V_q}{4}e^{i\Delta m\theta_q}\left(1+s_is_{i+1}e^{i(\phi_{m_{i+1}}-\phi_{m_i})}\right),\nonumber\\
e^{i\phi_{m}(k)}&=&C^A_m(k)/C^B_m(k),
\end{eqnarray}
where $\Delta m=m_i-m_{i+1}$ and $C^{A,B}_m(k)$ are the
coefficients of the Bloch components on A and B sub-lattices in
the electronic wave functions. Since one is mostly interested in
the properties of electronic states near the Fermi level, nearly
degenerate perturbation theory will be appropriate as long as the
perturbation is small $\delta=V_0/E_{10}\ll 1$, where
$E_{10}=v_R/R$ is the characteristic energy spacing between
subbands. Substituting the allowed transitions of
Eq.~(\ref{allowed transition}) into Eq.~(\ref{off-diagonal}) and
summing up, we can get the analytical expression for the band gap
at $k=k_F$:
\begin{eqnarray}
  E_g&\approx&2|H^{(3)}_{\pi\pi^*}|=\frac{V_1^2V_2}{2E^2_{10}}\sin(2\Delta\theta_{12})\left(\sin\frac{\pi}{3n}+\sin\frac{\pi}{6n}\right)\nonumber\\
&\approx&\frac{\pi}{4n}E_{10}\left(\frac{V_1^2V_2}{E^3_{10}}\right)\sin(2\Delta\theta_{12})\sim\delta^3R^{-2},
\end{eqnarray}
where $\Delta\theta_{12}=\theta_2-\theta_1$ and the dimensionless
potential $\delta$ is defined as $(V_1^2V_2)^{1/3}/E_{10}$.  The
$R^{-2}$ dependence makes $E_g$ of the order of a secondary band
gap in quasi-metallic SWNTs, which decays with the inverse square
of the radius. One should notice that $E_g$ is only related to the
relative angle $\Delta\theta_{12}$. When $\Delta\theta_{12}=0$,
the mirror planes of the two components coincide and perturbation
theory predicts no band gap whether or not the total potential
breaks the mirror symmetry. The maximum band gap happens at
$\Delta\theta_{12}=\pm \pi/4$, which corresponds to the configuration
that the nodes of the two potential components overlap, as
confirmed by numerical TB results when $n\ge5$
(Fig.~\ref{fig3}). Since this combination ($q_1=1$ and
$q_2=2$) always gives a secondary band gap unless $\Delta
\theta_{12}=0$, which can be tens or hundreds of meV for SWNTs
with a moderate radius, it may be an effective mechanism to induce
MST in armchair nanotubes and put them into tunable metallic
field-effect transistors~\cite{ROTK2004}.

In conclusion, we derived the selection rules for the
metal-semiconductor transition of armchair SWNTs under an external
circumferential perturbation within the orthogonal $\pi$-orbital TB
scheme. We evaluated the band gap opening as a function of the
external potential strength and its angular alignment with SWNT mirror
planes. Combinations of perturbations of different angular modes are
shown to open the gap up to 0.1 eV for $(6,6)$ SWNT and may represent
an effective mechanism of the metal-semiconductor transition.

This work was supported by the ARMY DURINT contract SIT 527826/222708 
and NSF grant ITR/SY 0121616. S.V.R acknowledges support
 through DoE grant DE-FG02-01ER45932, NSF grant ECS 04-03489
and EEC-0228390.

\bibliography{ref}
\end{document}